\documentclass{article}

\usepackage{arxiv}

\usepackage[utf8]{inputenc} 
\usepackage[T1]{fontenc}    
\usepackage[numbers,sort&compress]{natbib}
\usepackage{hyperref}       
\usepackage{url}            
\usepackage{booktabs}       
\usepackage{amsfonts}       
\usepackage{nicefrac}       
\usepackage{microtype}      
\usepackage{cleveref}       
\usepackage{lipsum}         
\usepackage{graphicx}
\usepackage{doi}
\usepackage{here}
\usepackage{algorithm}
\usepackage{algorithmicx,algpseudocode}
\usepackage{makecell}
\usepackage{tikz}
\usetikzlibrary{arrows.meta,positioning,calc,backgrounds,fit,shapes.geometric}
\title{Impact of File-Open Hook Points on Backup Ratio in ROFBS on XFS}

\newif\ifuniqueAffiliation
\uniqueAffiliationtrue

\ifuniqueAffiliation 
\author{ \href{https://orcid.org/0009-0001-0594-4620}{\includegraphics[scale=0.06]{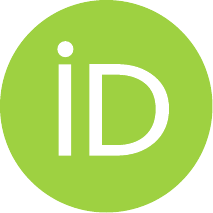}\hspace{1mm}Kosuke Higuchi} \\
	Kogakuin University\\ 1-24-2 Nishi-Shinjuku Shinjuku-ku\\ Tokyo, Japan\\
	\texttt{ed25001@ns.kogakuin.ac.jp} \\
	\And
	\href{https://orcid.org/0000-0001-5956-3455}{\includegraphics[scale=0.06]{orcid.pdf}\hspace{1mm}Ryotaro Kobayashi} \\
	Kogakuin University\\ 1-24-2 Nishi-Shinjuku Shinjuku-ku\\ Tokyo, Japan\\
	\texttt{ryo.kobayashi@cc.kogakuin.ac.jp} \\
}
\else
\usepackage{authblk}

\newbox{\orcid}\sbox{\orcid}{\includegraphics[scale=0.06]{orcid.pdf}} 
\author[1]{%
	\href{https://orcid.org/0000-0000-0000-0000}{\usebox{\orcid}\hspace{1mm}David S.~Hippocampus\thanks{\texttt{hippo@cs.cranberry-lemon.edu}}}%
}
\author[1,2]{%
	\href{https://orcid.org/0000-0000-0000-0000}{\usebox{\orcid}\hspace{1mm}Elias D.~Striatum\thanks{\texttt{stariate@ee.mount-sheikh.edu}}}%
}
\affil[1]{Department of Computer Science, Cranberry-Lemon University, Pittsburgh, PA 15213}
\affil[2]{Department of Electrical Engineering, Mount-Sheikh University, Santa Narimana, Levand}
\fi

\renewcommand{\shorttitle}{Impact of File-Open Hook Points on Backup Ratio in ROFBS on XFS}

\hypersetup{
pdftitle={Impact of File-Open Hook Points on Backup Ratio in ROFBS on XFS},
pdfsubject={cs.CR},
pdfauthor={Kosuke Higuchi, Ryotaro Kobayashi},
pdfkeywords={Ransomware defense, real-time backup, Linux kernel, eBPF, hook-point selection, file-open path, XFS},
}
\begin{document}
\maketitle

\begin{abstract}
Ransomware continues encrypting files during the delay between attack onset and detection.
ROFBS mitigates this problem by backing up pre-modification files in real time upon file-open events.
However, because the Linux file-open path traverses multiple kernel functions, it remains unclear how the choice of hook point affects defense effectiveness.

In this study, we kept the ROFBS mechanism fixed and changed only the hook points on the Linux file-open path.
We compared \texttt{may\_open}, \texttt{inode\_permission}, \texttt{do\_dentry\_open}, \texttt{security\_file\_open}, and \texttt{xfs\_file\_open} on AlmaLinux with XFS using three ransomware families: AvosLocker, Conti, and IceFire.
We used Backup Ratio as the main metric and also compared the number of encrypted files with backups and the total number of encrypted files.

The results showed that hook-point selection substantially affected both recoverability and damage scale.
For AvosLocker, \texttt{security\_file\_open} achieved the highest Backup Ratio (82.5\%).
For Conti and IceFire, \texttt{xfs\_file\_open} achieved the highest values (100.0\% and 63.2\%, respectively).
Moreover, \texttt{xfs\_file\_open} minimized the total number of encrypted files for all three ransomware families.

These results indicate that, in ROFBS, the layer at which file-open events are observed is a key design factor.
In particular, on XFS, hooking the filesystem-specific callback \texttt{xfs\_file\_open} may be advantageous when the goal is to reduce overall damage.
\end{abstract}


\section{Introduction}
Ransomware can encrypt or destroy a large number of files within a short period of time, severely affecting users' business continuity.
Accordingly, recent studies have emphasized not only accurate attack detection but also the extent to which pre-modification data can be preserved while an attack is in progress.
In practical deployments, the effectiveness of a defense mechanism depends not only on whether detection succeeds, but also on how many files ultimately remain recoverable.

To address this challenge, in our previous work we proposed ROFBS, a Real-Time Open-File Backup System that backs up pre-modification data in real time upon file-open events \cite{Higuchi2025}.
ROFBS aims to ensure high recoverability against ransomware-induced encryption by obtaining backups before file contents are modified.
Compared with approaches that attempt recovery only after modification has occurred, this approach is less susceptible to detection delay.

However, file-open processing in the Linux kernel does not complete within a single function; rather, it traverses multiple functions spanning the VFS layer, the LSM layer, and filesystem-specific processing.
For example, the open path includes \texttt{may\_open}, \texttt{inode\_permission} called within it, \texttt{do\_dentry\_open}, \texttt{security\_file\_open}, and filesystem-specific open functions.
Although all of these functions are related to file-open processing, they differ in the objects that can be observed, the timing of invocation, and the information available at each point.
Therefore, even without changing the ROFBS mechanism itself, the choice of kernel function used as the hook point may alter the timing at which backup creation becomes possible and the path information that can be obtained, thereby affecting both recoverability and the scale of encryption damage.

Existing studies have discussed the overall ROFBS architecture and its integration with detectors, but have not sufficiently isolated and evaluated which file-open-related kernel function should be used as the hook point \cite{Higuchi2025,HiguchiarXiv2025}.
This motivates an experiment in which the backup mechanism is fixed and only the hook point is treated as the independent variable.

In this study, we keep the basic ROFBS mechanism unchanged and evaluate how differences in hook point affect defense effectiveness in an XFS environment.
Specifically, we compare the behavior of \texttt{may\_open}, \texttt{inode\_permission}, \texttt{do\_dentry\_open}, \texttt{security\_file\_open}, and \texttt{xfs\_file\_open} when they are hooked using extended Berkeley Packet Filter (eBPF).
As the main metric, we use Backup Ratio, defined as the proportion of encrypted files for which a backup had been created in advance, and we also compare the number of encrypted files with backups and the total number of encrypted files.

The objective of this study is to identify the hook point best suited to ROFBS by evaluating how differences among file-open-related hook points affect Backup Ratio, the number of encrypted files with backups, and the total number of encrypted files, while keeping the ROFBS mechanism fixed in an XFS environment.
This study highlights the importance of hook-point selection in Linux ransomware defense and provides useful insights for the design of real-time backup mechanisms.

\section{Related Work}

Existing studies can be broadly categorized into three groups: research aimed at ransomware detection, research that mitigates damage using decoys or honeypots, and research that reduces damage through backup mechanisms.

First, among studies on ransomware detection, Zhuravchak et al. proposed a real-time ransomware detection system that integrates eBPF, machine learning (ML), and natural language processing (NLP) \cite{Danyil2023}.
Their system combines efficient data collection with eBPF, anomaly detection with ML, and text analysis with NLP, thereby achieving high detection accuracy.
Medhat et al. proposed a hybrid approach for detecting packed ransomware samples by scanning process memory dumps and dropped executables \cite{Mehnaz2018}.
By extending the YARA rule framework so that common ransomware characteristics can be described, their method achieved a high detection rate.
Kok et al. proposed a machine-learning-based pre-encryption detection algorithm that detects ransomware with high accuracy before encryption begins \cite{kok2019}.
Song et al. proposed an anomaly detection method for Android platforms based on resource-related indicators such as CPU usage, memory consumption, and I/O rate \cite{Song2016}.
In addition, Cusack et al. built a stream processor on a Programmable Forwarding Engine (PFE) and proposed a method for identifying malicious network behavior using a random forest classifier, demonstrating the possibility of pre-encryption detection through flow-based fingerprinting \cite{Greg2018}.

From the perspective of monitoring infrastructure itself, Caviglione et al. proposed a method that leverages eBPF to efficiently trace and monitor software process behavior \cite{Luca2021}.
Although this work is not specific to ransomware, it showed that filesystem changes and hidden communications can be observed with low overhead, thereby supporting the effectiveness of using eBPF to capture security-relevant events on Linux.

As for approaches that aim to stop damage in progress, G\'omez-Hern\'andez et al. proposed R-Locker, which defends against ransomware by monitoring accesses to decoy files \cite{gomez2018}.
Their method prevents the progression of encryption by stopping ransomware once access to a decoy file is detected.
Similarly, Zhuravchak et al. proposed a method that mitigates filesystem damage using honeypot techniques and symbolic links \cite{Zhuravchak2021}.
Lee et al. focused on the fact that ransomware binaries often whitelist specific file extensions, and proposed a defense method that protects important files by randomly changing their extensions \cite{Lee2019}.

There have also been studies that reduce damage through backup mechanisms.
Fujinoki et al. proposed PDPZR, a backup system that creates a backup whenever data are updated and manages redundancy by deleting old versions \cite{fujinoki2023}.
Gujar et al. proposed an automatic backup system that detects newly added files in real time and immediately backs them up to a dedicated folder on an SSD \cite{Gujar2023}.
Oujezsky et al. proposed the Intelligent Malware Defense System (IMDS), which uses AI and hash functions to verify file integrity before backup and blocks the backup process when anomalies are detected \cite{Oujezsky2023}.

In summary, prior work has explored improvements in detection accuracy, damage suppression using decoys and honeypots, and damage mitigation through backup.
However, for a mechanism such as ROFBS, which preserves pre-modification data upon file-open events, there has been insufficient evaluation of which hook point in the Linux file-open path should be selected.
Our work differs in that it keeps the ROFBS mechanism itself fixed while changing only the hook point, and compares the resulting effects on Backup Ratio, the number of encrypted files with backups, and the total number of encrypted files.

\section{Background}
This section introduces ROFBS, which is used in the experiments, the eBPF and BCC technologies used in its implementation, and an overview of the open-related kernel functions evaluated in this study.

\subsection{Overview of ROFBS}
ROFBS is a real-time file backup system proposed by the authors to mitigate ransomware damage \cite{Higuchi2025}.
Previous ransomware defense studies based on machine learning have mainly focused on improving detection methods and detection accuracy.
In practice, however, file encryption continues even during the period between the start of ransomware execution and its detection.
Therefore, improving detection performance alone is not sufficient to fully suppress encryption damage itself.

Kok et al. pointed out that, although many studies adopt machine learning for ransomware detection, relatively few propose models that directly prevent or mitigate the attack itself \cite{kok2019review}.
This suggests that, in order to cope with increasingly sophisticated and diverse ransomware attacks, it is important to design defense models aimed not only at detection but also at damage mitigation.
Moreover, despite the increase in ransomware targeting Linux, defense models and countermeasures for Linux are still not sufficiently developed \cite{trendmicro2022,Surati2017}.
Whereas many endpoint security products are available in Windows environments, the options for Linux remain limited, highlighting the need for stronger defense techniques.

Against this background, the authors first proposed a model incorporating a file-protection mechanism to mitigate encryption damage occurring during the detection delay \cite{candar2023}, and then developed ROFBS as its extension \cite{Higuchi2025}.
ROFBS captures file-related events using eBPF and creates backups before files are modified, thereby preserving pre-modification data even while ransomware is running.
In this way, ROFBS aims to mitigate ransomware damage in Linux environments while combining real-time monitoring with immediate response.

In the ROFBS implementation, file paths are received incrementally from the kernel in reverse order, and complete paths are reconstructed by combining multiple events.
In addition, backup files are assigned the \texttt{.tmp} extension.
This is because previous work has shown that ransomware tends to exclude files with certain extensions, such as \texttt{.tmp} and \texttt{.exe}, from encryption targets \cite{saleh2022}.
After the malicious process has been stopped, the original file can be restored by renaming the backup file back to its original extension when necessary.

Algorithm~\ref{algo:backup} shows the backup procedure used in ROFBS.

\makeatletter
\renewcommand{\ALG@name}{Algorithm}
\makeatother
\begin{algorithm}[H]
  \caption{Backup procedure of ROFBS}
  \label{algo:backup}
  \begin{algorithmic}[1]
    \Require File path ($file\_path$) and list of protected directories ($protected\_directories$)
    \Function{CreateBackupFile}{$file\_path$, $protected\_directories$}
      \If{$file\_path$ is under one of $protected\_directories$ and has not yet been backed up}
        \State Save the pre-modification contents of $file\_path$ as a backup file with the \texttt{.tmp} extension.
        \State Record $file\_path$ as already backed up.
      \EndIf
    \EndFunction
    \Function{RestoreBackup}{$file\_path$, $is\_malicious$}
      \If{$is\_malicious$ is true and a corresponding \texttt{.tmp} backup exists for $file\_path$}
        \State Rename the \texttt{.tmp} backup to the original file name and restore $file\_path$.
      \EndIf
    \EndFunction
  \end{algorithmic}
\end{algorithm}

In the evaluation reported in \cite{Higuchi2025}, a backup ratio of 100\% was achieved for both Monti and Conti across all machine-learning algorithms evaluated.
High backup ratios were also reported for other ransomware families.
These results suggest that ROFBS can be effective as a Linux-oriented ransomware defense mechanism.

\subsection{Extended Berkeley Packet Filter}
eBPF is a mechanism for safely and efficiently executing sandboxed programs in a privileged context of the Linux kernel \cite{eBPF}.
By using eBPF, it is possible to extend kernel capabilities for tracing, performance analysis, networking, and security monitoring without modifying kernel source code or loading kernel modules.
In addition, eBPF programs are checked by the verifier before being loaded into the kernel, and are executed only after their safety has been confirmed.
At runtime, JIT compilation can provide high execution efficiency.

In general, the functions available in eBPF and the hook points to which programs can be attached depend on the kernel version.
To address this issue, recent approaches such as BPF CO-RE, which relies on BTF and libbpf, have been developed to improve portability across different kernel versions.
However, the implementation in this study uses BCC, described below.

\subsection{BPF Compiler Collection}
The BPF Compiler Collection (BCC) is a toolkit for developing eBPF programs \cite{bcc}.
With BCC, kernel-side instrumentation is typically written in C, while user-space components for loading programs and processing collected data can be written in Python or Lua.
This makes it relatively easy to implement tracing and monitoring mechanisms based on eBPF.
BCC also provides a variety of sample tools and utilities, enabling developers to rapidly build eBPF-based tracing and monitoring tools.

In this study, we use probes to hook kernel functions.
Typical examples include kprobes, which fire at function entry, and kretprobes, which fire when a function returns.
Because the information available before and after function execution differs, it is important to choose the appropriate probe type depending on the purpose.

\subsection{File-Open Path in Linux}

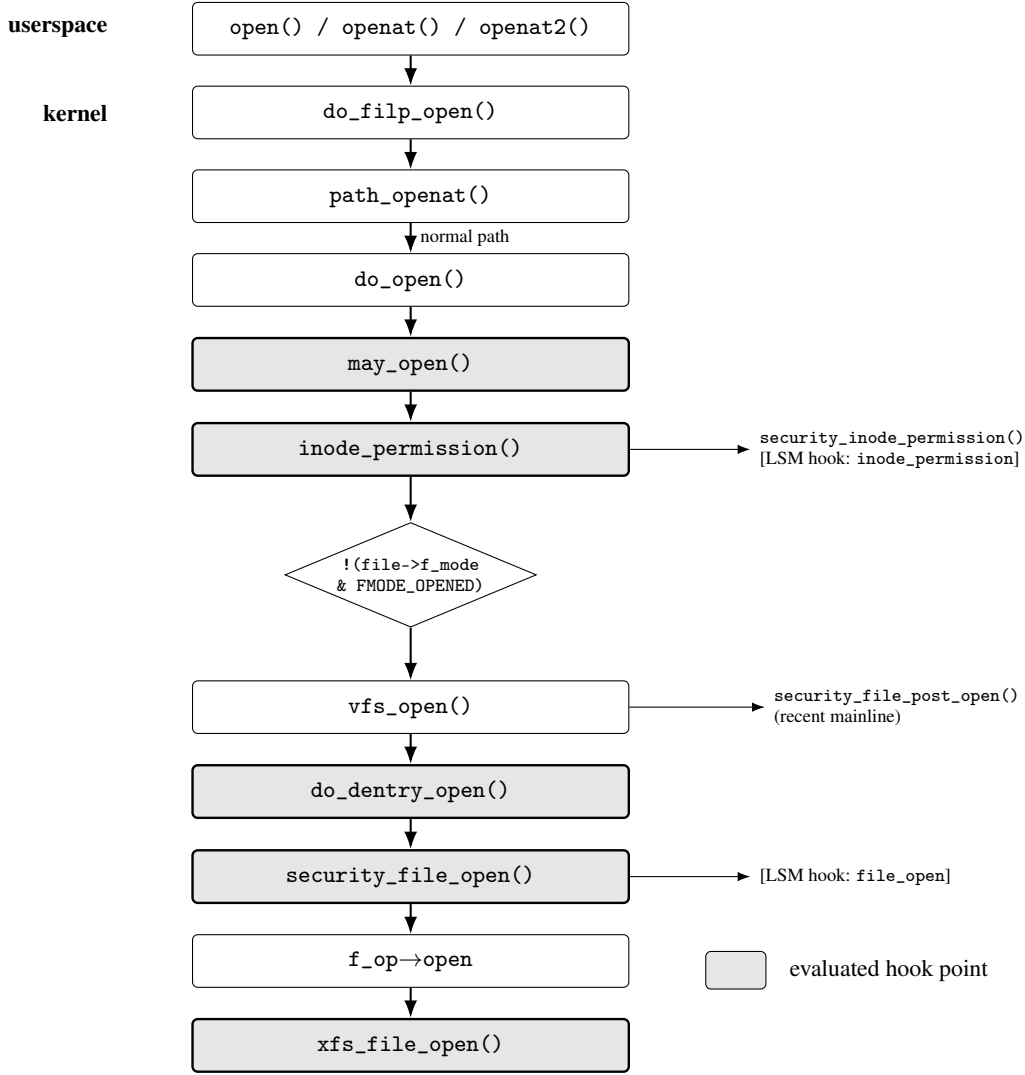
\begin{figure}[t]
    \centering
    \begin{tikzpicture}[
        >=Latex,
        font=\footnotesize,
        node distance=4mm and 12mm,
        fn/.style={
            draw,
            rounded corners=2pt,
            minimum width=5.8cm,
            minimum height=7mm,
            align=center,
            inner sep=2pt,
            font=\ttfamily\footnotesize
        },
        hook/.style={
            fn,
            fill=gray!20,
            line width=0.9pt
        },
        cond/.style={
            draw,
            diamond,
            aspect=2.4,
            align=center,
            inner sep=1pt,
            font=\ttfamily\scriptsize
        },
        ann/.style={
            font=\scriptsize,
            align=left
        },
        lab/.style={
            font=\footnotesize\bfseries,
            align=right
        }
    ]

    \node[fn]   (u)   {open() / openat() / openat2()};
    \node[fn, below=of u]   (f1)  {do\_filp\_open()};
    \node[fn, below=of f1]  (f2)  {path\_openat()};
    \node[fn, below=of f2]  (f3)  {do\_open()};
    \node[hook, below=of f3] (h1) {may\_open()};
    \node[hook, below=of h1] (h2) {inode\_permission()};
    \node[cond, below=6mm of h2] (c1) {!(file->f\_mode\\\& FMODE\_OPENED)};
    \node[fn,   below=7mm of c1] (f4) {vfs\_open()};
    \node[hook, below=of f4] (h3) {do\_dentry\_open()};
    \node[hook, below=of h3] (h4) {security\_file\_open()};
    \node[fn,   below=of h4] (f5) {f\_op$\rightarrow$open};
    \node[hook, below=of f5] (h5) {xfs\_file\_open()};

    \node[lab, left=10mm of u]  {userspace};
    \node[lab, left=10mm of f1] {kernel};

    \draw[->, thick] (u)  -- (f1);
    \draw[->, thick] (f1) -- (f2);
    \draw[->, thick] (f2) -- node[right, font=\scriptsize] {normal path} (f3);
    \draw[->, thick] (f3) -- (h1);
    \draw[->, thick] (h1) -- (h2);
    \draw[->, thick] (h2) -- (c1);
    \draw[->, thick] (c1) -- (f4);
    \draw[->, thick] (f4) -- (h3);
    \draw[->, thick] (h3) -- (h4);
    \draw[->, thick] (h4) -- (f5);
    \draw[->, thick] (f5) -- (h5);

    \node[ann, right=16mm of h2] (a1) {\texttt{security\_inode\_permission()}\\{}[LSM hook: \texttt{inode\_permission}]};
    \draw[->] (h2.east) -- ++(7mm,0) |- (a1.west);

    \node[ann, right=16mm of h4] (a2) {[LSM hook: \texttt{file\_open}]};
    \draw[->] (h4.east) -- ++(7mm,0) |- (a2.west);

    \node[ann, right=18mm of f4] (a3) {\texttt{security\_file\_post\_open()}\\(recent mainline)};
    \draw[->] (f4.east) -- ++(7mm,0) |- (a3.west);

    \node[
        draw,
        rounded corners=2pt,
        fill=gray!20,
        minimum width=8mm,
        minimum height=5mm,
        anchor=west
    ] at ($(h5.east)+(10mm,10mm)$) {};
    \node[anchor=west, font=\footnotesize]
        at ($(h5.east)+(20mm,10mm)$) {evaluated hook point};

    \end{tikzpicture}
    \caption{Simplified Linux open path relevant to this study.
    Gray boxes indicate the hook points evaluated in the experiments.
    Helper functions and special-case branches are omitted except for several annotated points.}
    \label{fig:open-path}
\end{figure}

In this study, multiple kernel functions along the Linux open path are compared as candidates for triggering ROFBS backups.
Figure~\ref{fig:open-path} shows an overview of the Linux open path relevant to this study.
The figure does not attempt to cover all helper functions or special-case branches in open processing; rather, it simplifies the main path in order to show the relative positions of the hook points evaluated in this study.
Table~\ref{tab:hookpoint-summary} summarizes the layer and the main information available at each hook point.

\begin{table}[H]
    \centering
    \caption{Layers and main information available at each hook point}
    \scalebox{0.90}{
    \begin{tabular}{p{3.4cm}p{3.2cm}p{8.2cm}}
        \hline
        \textbf{Hook point} & \textbf{Layer} & \textbf{Main information available} \\
        \hline
        \texttt{may\_open}
        & Early decision stage
        & \texttt{path}, \texttt{dentry}, \texttt{inode}, open flag \\

        \texttt{inode\_permission}
        & Permission-check stage
        & \texttt{inode}, access mask \\

        \texttt{do\_dentry\_open}
        & Common VFS open-execution stage
        & \texttt{file}, \texttt{f\_inode}, \texttt{f\_mapping}, \texttt{f\_op} \\

        \texttt{security\_file\_open}
        & Late common VFS stage
        & \texttt{file}, state at open time \\

        \texttt{xfs\_file\_open}
        & Filesystem-specific stage
        & \texttt{file}, \texttt{inode} \\
        \hline
    \end{tabular}}
    \label{tab:hookpoint-summary}
\end{table}

Below, the role of each hook point is described based on the open path in the target kernel version used in our evaluation.

In Linux, an open request issued from user space is processed through pathname lookup and open handling in the VFS.
During this process, path resolution, acquisition of dentries and inodes, access checks, and initialization of the file structure are performed in stages.
Simplifying the open path considered in this paper, \texttt{may\_open} performs permission checking through \texttt{inode\_permission}, after which control proceeds via \texttt{vfs\_open} to \texttt{do\_dentry\_open}, where \texttt{security\_file\_open} is executed before the filesystem-specific open callback.
Thus, Linux open processing is not completed within a single function, but consists of an early decision stage, a common VFS open-execution stage, and a filesystem-specific stage.

The first hook point considered in this study, \texttt{may\_open}, is a decision function located in a relatively early part of the open path.
In \texttt{may\_open}, the type of the target inode and the consistency of the open flag are checked, and access permission is then evaluated through \texttt{inode\_permission}.
Therefore, hooking \texttt{may\_open} means observing events before the actual open execution and before the filesystem-specific open callback is reached.
This position has the advantage of capturing open requests at an early stage, but it may also include events that do not proceed to later stages.

The second hook point, \texttt{inode\_permission}, is responsible for permission checking on an inode.
Although this function is invoked from \texttt{may\_open} on the open path, it is not specific to open operations.
Accordingly, when \texttt{inode\_permission} is used as a hook point, an implementation issue is how to distinguish invocations caused specifically by open.
Moreover, because this function primarily handles an inode and an access mask, and does not directly provide path- or file-based context, the contextual information available at this point is more limited than at other hook points.

The third hook point, \texttt{do\_dentry\_open}, is located at the common VFS open-execution stage.
At this point, major fields in the file structure, such as \texttt{f\_inode}, \texttt{f\_mapping}, and \texttt{f\_op}, have already been initialized.
Therefore, \texttt{do\_dentry\_open} reflects a more concrete open state than a simple permission check.
For a mechanism such as ROFBS, which performs backups on a per-file basis, this makes it a hook point at which the target file can be identified relatively clearly.

The fourth hook point, \texttt{security\_file\_open}, is executed inside \texttt{do\_dentry\_open} before the filesystem-specific open callback is invoked.
This function serves as the entry point for LSM-based state handling and additional checks at open time.
Therefore, using \texttt{security\_file\_open} allows events to be captured at a point close to the actual open operation while still remaining on the filesystem-independent common path.

The fifth hook point is a filesystem-specific open callback.
In this study, we evaluate \texttt{xfs\_file\_open}.
This is the callback for regular files, whereas directories use a separate open callback.
As a result, compared with hook points on the common VFS path, a filesystem-specific open callback may observe a more restricted set of events.
On the other hand, because it is located near the final stage in the filesystem-specific layer, it can observe open events in a manner that more closely reflects the implementation of the underlying filesystem.

In summary, although all hook points compared in this study are related to open processing, they are distributed across different layers: an early decision stage, a common VFS execution stage, and a filesystem-specific stage.
As a result, the type of information available, the range of open events observed, and the timing at which ROFBS starts backup processing may differ across hook points.
To evaluate how these differences affect the number of encrypted files with backups, the total number of encrypted files, and Backup Ratio, this study compares these five hook points \cite{linux_namei,linux_open,linux_xfs_file}.

\subsection{Experimental Setup}

The experimental environment was constructed using VirtualBox.
Table~\ref{tab:evaluation} summarizes the specifications of the host machine and the virtual machine.
Because this study evaluates \texttt{xfs\_file\_open}, the target directory \texttt{victim} in the virtual machine was placed on an XFS filesystem.

\begin{table}[H]
    \centering
    \caption{Experimental environment}
    \scalebox{0.90}{
    \begin{tabular}{lcc}
        \hline
        & \textbf{Host Machine} & \textbf{Virtual Machine} \\
        \hline
        CPU     & Intel Core i7-12650H & 1 vCPU \\
        Memory  & 32 GB (DDR4 2400 MT/s) & 8 GB \\
        Storage & 512 GB (M.2 NVMe SSD) & 128 GB (virtual disk) \\
        OS      & Windows 11 & AlmaLinux 10.1 \\
        \hline
    \end{tabular}}
    \label{tab:evaluation}
\end{table}

The experimental settings follow the basic policy of prior work, while being adapted to the environment used in this study \cite{Higuchi2025,HiguchiarXiv2025}.
Specifically, we created a \texttt{victim} directory containing 4,385 files taken from the \texttt{tiny} folder of the NapierOne dataset \cite{Napierone2022}.
The virtual environment included this \texttt{victim} directory, together with the files and libraries required to run the detection model and ROFBS.

Three ransomware samples were used in this study.
These samples were obtained between May and October 2023.
Each sample was identified by its SHA256 hash, and its name follows the signature name assigned at the time of acquisition.

In principle, each sample was executed with default settings, and the \texttt{--path} argument was provided only when specifying the target path was necessary.
That is, we used only the arguments required for execution or explicitly listed in the help menu, and did not specify additional options such as thread count or background execution.
IceFire does not support user-defined encryption targets, and therefore this argument was not applied to IceFire.

\begin{table}[H]
    \centering
    \caption{Ransomware samples used in the experiments}
    \scalebox{0.90}{
    \begin{tabular}{lc}
        \hline
        \textbf{Ransomware} & \textbf{SHA256} \\
        \hline
        IceFire    & e9cc7fdfa3cf40ff9c3db0248a79f4817b170f2660aa2b2ed6c551eae1c38e0b \\
        AvosLocker & 0cd7b6ea8857ce827180342a1c955e79c3336a6cf2000244e5cfd4279c5fc1b6 \\
        Conti      & 95776f31cbcac08eb3f3e9235d07513a6d7a6bf9f1b7f3d400b2cf0afdb088a7 \\
        \hline
    \end{tabular}}
    \label{tab:ransomware}
\end{table}

Table~\ref{tab:ml-algo} lists the machine-learning algorithm adopted in this study.
Accuracy and FPR were evaluated using logs collected during the execution of AvosLocker.
These logs were unparsed and had the same format as those collected in this study.
Each log entry was manually labeled as benign or malicious.
Accuracy was calculated as the agreement rate between the model predictions and the manual labels.
FPR was calculated as the proportion of benign logs incorrectly classified as malicious.
Based on the Accuracy and FPR results, we adopted Random Forest (RF).

\begin{table}[H]
    \centering
    \caption{Machine-learning algorithm adopted in this study}
    \scalebox{0.90}{
        \begin{tabular}{l r r}
            \hline
            \textbf{Algorithm} & \textbf{Accuracy (\%)} & \textbf{FPR} \\
            \hline
            Random Forest (RF) & 97.2 & 0.0068 \\
            \hline
        \end{tabular}}
    \label{tab:ml-algo}
\end{table}

\subsection{Backup Ratio}

To evaluate the effectiveness of the backup mechanism, this study uses Backup Ratio as an evaluation metric.
Backup Ratio represents the proportion of encrypted files for which backups had been created in advance, and is defined as follows:

\begin{equation}
  \mathrm{Backup\ Ratio}[\%] =
  \frac{N_{\mathrm{backup}}}{N_{\mathrm{encrypted}}} \times 100
  \label{eq:backup-ratio}
\end{equation}

Here, $N_{\mathrm{backup}}$ denotes the number of encrypted files for which pre-created backups existed, and $N_{\mathrm{encrypted}}$ denotes the total number of encrypted files.

ROFBS was proposed to mitigate damage caused during the delay before ransomware is detected.
Accordingly, this study focuses not on the number of files that were not encrypted, but on how many encrypted files remained recoverable.
At the same time, because Backup Ratio depends on the denominator $N_{\mathrm{encrypted}}$, the magnitude of change in the ratio is not uniform when $N_{\mathrm{encrypted}}$ differs substantially across hook points, even if the absolute difference is the same.
Therefore, while Backup Ratio is used as the primary metric in this study, its interpretation is supplemented by the absolute values of the number of encrypted files with backups ($B$) and the total number of encrypted files ($E$).
\subsection{Results}

Table~\ref{tab:hook_counts} shows, for each hook point, the number of encrypted files with backups ($B$) and the total number of encrypted files ($E$).

\begin{table}[H]
    \centering
    \caption{Numbers of encrypted files with backups ($B$) and total encrypted files ($E$) for each hook point}
    \scalebox{0.90}{
    \begin{tabular}{lcccccc}
        \hline
        & \multicolumn{2}{c}{\textbf{AvosLocker}} & \multicolumn{2}{c}{\textbf{Conti}} & \multicolumn{2}{c}{\textbf{IceFire}} \\
        \textbf{Hook Point} & \textbf{$B$} & \textbf{$E$} & \textbf{$B$} & \textbf{$E$} & \textbf{$B$} & \textbf{$E$} \\
        \hline
        \texttt{may\_open}            & 9  & 32 & 1  & 46 & 2  & 30 \\
        \texttt{inode\_permission}    & 7  & 15 & 1  & 46 & 9  & 37 \\
        \texttt{do\_dentry\_open}     & 21 & 29 & 1  & 46 & 9  & 35 \\
        \texttt{security\_file\_open} & 33 & 40 & 1  & 46 & 9  & 36 \\
        \texttt{xfs\_file\_open}      & 3  & 7  & 18 & 18 & 12 & 19 \\
        \hline
    \end{tabular}}
    \label{tab:hook_counts}
\end{table}

As shown in Table~\ref{tab:hook_counts}, the choice of hook point affects both $B$ and $E$.
For AvosLocker, \texttt{security\_file\_open} yielded the largest number of encrypted files with backups, recording $B=33$.
In contrast, although \texttt{xfs\_file\_open} yielded only $B=3$, it minimized the total number of encrypted files to $E=7$, showing a different trend from the other hook points.
For Conti, \texttt{may\_open}, \texttt{inode\_permission}, \texttt{do\_dentry\_open}, and \texttt{security\_file\_open} all produced the same result, namely $B=1, E=46$.
By contrast, \texttt{xfs\_file\_open} yielded $B=18, E=18$, showing a markedly different distribution from the other four hook points.
For IceFire as well, \texttt{xfs\_file\_open} yielded $B=12, E=19$, which was the smallest value of $E$ among all hook points.

Table~\ref{tab:hook_ratio} shows the Backup Ratio values calculated according to Eq.~(\ref{eq:backup-ratio}).

\begin{table}[H]
    \centering
    \caption{Backup Ratio (\%) for each hook point}
    \scalebox{0.90}{
    \begin{tabular}{lccc}
        \hline
        \textbf{Hook Point} & \textbf{AvosLocker} & \textbf{Conti} & \textbf{IceFire} \\
        \hline
        \texttt{may\_open}            & 28.1 & 2.2   & 6.7  \\
        \texttt{inode\_permission}    & 46.7 & 2.2   & 24.3 \\
        \texttt{do\_dentry\_open}     & 72.4 & 2.2   & 25.7 \\
        \texttt{security\_file\_open} & 82.5 & 2.2   & 25.0 \\
        \texttt{xfs\_file\_open}      & 42.9 & 100.0 & 63.2 \\
        \hline
    \end{tabular}}
    \label{tab:hook_ratio}
\end{table}

Table~\ref{tab:hook_ratio} shows that, for AvosLocker, \texttt{security\_file\_open} achieved the highest Backup Ratio at 82.5\%, followed by \texttt{do\_dentry\_open} at 72.4\%.
For Conti, \texttt{xfs\_file\_open} reached 100.0\%, substantially outperforming the other hook points.
For IceFire, \texttt{xfs\_file\_open} also achieved the highest value, at 63.2\%.
These results indicate that \texttt{security\_file\_open} yielded the highest Backup Ratio for AvosLocker, whereas \texttt{xfs\_file\_open} achieved the highest Backup Ratio for Conti and IceFire.
By contrast, \texttt{may\_open} remained relatively low for all three ransomware families.
However, because Backup Ratio uses the total number of encrypted files as its denominator, it should be interpreted together with the absolute values of $B$ and $E$ shown in Table~\ref{tab:hook_counts}.
\section{Discussion}

In this section, let $B$ denote the number of encrypted files with backups and $E$ denote the total number of encrypted files.
In this study, the ROFBS mechanism itself was kept fixed, and only the hook point was changed, yet substantial differences were observed in both $B$ and $E$.
This indicates that changing the hook point is not merely an implementation detail, but rather changes the range and context of the open events observed by ROFBS.

Although Backup Ratio is a useful normalized metric, the denominator $E$ varies across hook points, which means that defense effectiveness cannot be determined solely from the ratio itself.
In particular, when $E$ is small, even small differences in $B$ or $E$ can lead to large changes in Backup Ratio.
Accordingly, the following discussion interprets not only Backup Ratio but also the absolute values of $B$ and $E$.

\subsection{Effect of Hook Position}

All hook points compared in this study reside in kernel space, but their relative positions differ.
\texttt{may\_open} and \texttt{inode\_permission} are located in the early part of the VFS path and are invoked relatively close to the open request issued from user space.
By contrast, \texttt{do\_dentry\_open} and \texttt{security\_file\_open} are located in the later part of the common VFS open-execution path.
Furthermore, \texttt{xfs\_file\_open} is a filesystem-specific open callback for regular files in XFS, and among the candidates considered here it is the closest to the filesystem-specific layer \cite{linux_namei,linux_open,linux_xfs_file}.

These differences directly affect both the information available to ROFBS and the range of events that can be observed.
Because \texttt{may\_open} is an early decision function and performs permission checking through \texttt{inode\_permission}, upper-level hooks may observe a broader range of events rather than only files that actually proceed to the open-execution stage \cite{linux_namei}.
In particular, \texttt{inode\_permission} mainly handles an inode and an access mask, and therefore provides more limited contextual information for file identification than later hooks that can directly access \texttt{struct file}.
This interpretation is consistent with the results for \texttt{may\_open} and \texttt{inode\_permission}, where $B$ tended to remain small and $E$ was not sufficiently suppressed.

By contrast, \texttt{do\_dentry\_open} and \texttt{security\_file\_open} can use \texttt{struct file} at a stage closer to the actual open execution, allowing ROFBS to identify target files more concretely.
Experimentally, these later hooks increased $B$ compared with \texttt{may\_open} and \texttt{inode\_permission}.
In particular, \texttt{security\_file\_open} recorded a relatively large number of encrypted files with backups overall.
However, this did not translate into suppression of $E$, suggesting that, at least under the present experimental conditions, increasing the number of backups alone is not sufficient to fully characterize protection effectiveness.

\subsection{Why \texttt{xfs\_file\_open} Was Advantageous in This Setting}

\texttt{xfs\_file\_open} is the lowest-level hook point evaluated in this study and is specific to the filesystem layer.
In XFS, the open callback for regular files is separated from that for directories, and \texttt{xfs\_file\_open} is invoked only for regular files \cite{linux_xfs_file}.
As a result, compared with hooks placed on the common VFS path, ROFBS performs backup processing on a more restricted set of targets.

In the present results, \texttt{xfs\_file\_open} did not always maximize $B$, but it minimized $E$ for all three ransomware families.
This suggests that what matters for ROFBS is not simply increasing the number of observed events, but reacting preferentially to open events that are more likely to lead to actual encryption.
Whereas the common VFS path can observe a broad set of open events, some of them may be of relatively low priority from the perspective of ROFBS.
By contrast, because \texttt{xfs\_file\_open} restricts observation to regular-file opens, it may have enabled backup-processing resources to be used more effectively.

The fact that \texttt{security\_file\_open} increased $B$ but did not sufficiently suppress $E$ also suggests that broadly observing open events on the common path is not enough.
In other words, the effectiveness of ROFBS depends not only on how deep the observation point is, but also on how well that point filters out unnecessary open events.
Under the present XFS-based setting, the filesystem-specific callback \texttt{xfs\_file\_open} may therefore have been advantageous in this respect.

\subsection{Differences Across Ransomware Families}

Looking at differences across ransomware families, for AvosLocker, \texttt{security\_file\_open} recorded the largest number of encrypted files with backups and also achieved the highest Backup Ratio.
At the same time, the smallest value of $E$ was obtained with \texttt{xfs\_file\_open}.
This indicates that, for AvosLocker, the hook point that is favorable for maximizing recoverability does not necessarily coincide with the one that is favorable for minimizing the overall scale of encryption damage.
Therefore, for AvosLocker, it is not appropriate to conclude that \texttt{security\_file\_open} is the best choice solely on the basis of Backup Ratio; both $B$ and $E$ must also be considered.

For Conti, \texttt{may\_open}, \texttt{inode\_permission}, \texttt{do\_dentry\_open}, and \texttt{security\_file\_open} all produced identical results, whereas only \texttt{xfs\_file\_open} showed a marked improvement.
This suggests that, at least under the present experimental conditions, lower-level filesystem-specific hooks may be more effective than upper VFS hooks or common-path hooks for ransomware with behavior similar to Conti.

For IceFire as well, \texttt{xfs\_file\_open} showed favorable results in terms of both $B$ and $E$, supporting the same tendency.
Thus, although the behavior varies across ransomware families, the present experiments suggest that, on XFS, using a filesystem-specific open callback is advantageous particularly for suppressing $E$.

\subsection{Implications for ROFBS Design}

Overall, these results indicate that, in ROFBS, hook-point selection affects not only Backup Ratio but also the overall scale of encryption damage.
Upper-level VFS hooks can broadly observe open requests, but were less likely to translate into effective backups for ROFBS.
Later hook points on the common VFS path increased the number of encrypted files with backups, but had limited ability to suppress the total number of encrypted files.
By contrast, the XFS filesystem-specific open callback \texttt{xfs\_file\_open} consistently yielded smaller values of $E$ under the present experimental conditions by concentrating observation on regular-file opens.

Therefore, at least when ROFBS is deployed on XFS, hooking at a point close to the filesystem-specific open path may be more effective than hooking upper VFS or common open-path functions.
These results indicate that the layer at which file-open events are observed is a fundamental design factor in real-time backup mechanisms.

\subsection{Limitations}

This study has several limitations.
First, the filesystem-specific hook evaluated here is \texttt{xfs\_file\_open} in XFS, and the findings cannot be directly generalized to other filesystems.
For example, ext4 implements \texttt{ext4\_file\_open} as the regular-file open callback, and its internal processing at open time differs from that of XFS \cite{linux_ext4_file,linux_xfs_file}.

Second, this study focuses primarily on effectiveness as evaluated through Backup Ratio and file counts, and does not provide a detailed evaluation of performance, CPU utilization, event volume, or queue behavior in user space.
Accordingly, further investigation is needed to clarify how the observed differences relate to backup-processing resource consumption and event concentration.

Third, the evaluation in this study is based on XFS, the three ransomware families examined here, and a single workload.
Whether the same tendencies hold for other filesystems, other ransomware families, or different workload conditions remains future work.

Fourth, for hook points such as \texttt{inode\_permission}, where path information and \texttt{struct file} cannot be obtained directly, paths are reconstructed using representative dentries.
This reconstruction method itself may therefore affect the results, and should be kept in mind when interpreting comparisons across hook points.
\section{Conclusion}

In this study, we evaluated how changing only the hook point, while keeping the ROFBS mechanism itself unchanged, affected the number of encrypted files with backups, the total number of encrypted files, and Backup Ratio.
As comparison targets, we used five hook points: \texttt{may\_open}, \texttt{inode\_permission}, \texttt{do\_dentry\_open}, \texttt{security\_file\_open}, and \texttt{xfs\_file\_open}, and conducted experiments using AvosLocker, Conti, and IceFire.

The experimental results showed that the choice of hook point has a substantial impact on the effectiveness of ROFBS.
For AvosLocker, \texttt{security\_file\_open} achieved the highest Backup Ratio, whereas for Conti and IceFire, \texttt{xfs\_file\_open} achieved the highest Backup Ratio.
Moreover, \texttt{xfs\_file\_open} minimized the total number of encrypted files for all three ransomware families, indicating an overall advantage in suppressing the scale of damage.
By contrast, \texttt{may\_open} remained relatively low in Backup Ratio across all three ransomware families.

However, because Backup Ratio uses the total number of encrypted files as its denominator, defense effectiveness cannot be determined solely from this ratio.
In particular, for AvosLocker, \texttt{security\_file\_open} achieved the highest Backup Ratio, while \texttt{xfs\_file\_open} minimized the total number of encrypted files.
This indicates that the hook point favorable for maximizing recoverability does not necessarily coincide with the one favorable for minimizing the total number of encrypted files.
Therefore, when evaluating ROFBS, it is important to interpret Backup Ratio together with the absolute values of the number of encrypted files with backups and the total number of encrypted files.

These results indicate that what matters in ROFBS is not merely whether file-open events are monitored, but which layer of the open path is observed.
Differences among an upper VFS decision stage, a common VFS execution stage, and a filesystem-specific open callback affect the nature of the observed events, the information obtainable at the hook point, and the timing at which backup creation begins, and these differences ultimately appear as differences in defense effectiveness.
Accordingly, in a real-time backup mechanism such as ROFBS, the choice of hook point is itself a fundamental design factor.

The results of this study suggest that, at least when ROFBS is deployed on XFS, hooking near a filesystem-specific open callback such as \texttt{xfs\_file\_open} can be advantageous from the perspective of suppressing the scale of damage.
At the same time, this study focuses on evaluation on XFS, and the findings cannot be directly generalized to other filesystems.
Future work includes evaluation on other filesystems such as ext4, more detailed analyses including performance and event volume, and validation of reproducibility across other kernel lines and execution environments.

Overall, this study demonstrates that hook-point selection in ROFBS affects not only Backup Ratio but also the total number of encrypted files, and provides useful insights for the design of Linux-oriented ransomware defense mechanisms.
\bibliographystyle{unsrt}
\bibliography{open}

@inproceedings{kok2019review,
  title={Ransomware , Threat and Detection Techniques : A Review},
  author={S. H. Kok and Azween Bin Abdullah and Noor Zaman Jhanjhi and Mahadevan Supramaniam},
  booktitle={International Journal of Computer Science and Network Security},
  year={2019},
  volume = {19},
  number = {2},
  pages = {136--146},
}

@misc{HiguchiarXiv2025,
  author        = {Kosuke Higuchi and Ryotaro Kobayashi},
  title         = {{ROFBS}$\alpha$: Real Time Backup System Decoupled from ML Based Ransomware Detection},
  year          = {2025},
  eprint        = {2504.14162},
  archivePrefix = {arXiv},
  primaryClass  = {cs.CR},
  howpublished  = {\url{https://arxiv.org/abs/2504.14162}},
  note          = {arXiv:2504.14162}
}

@article{Higuchi2025,
  author    = {Kosuke Higuchi and Ryotaro Kobayashi},
  title     = {Real-time open-file backup system with machine-learning detection model for ransomware},
  journal   = {International Journal of Information Security},
  volume    = {24},
  number    = {1},
  pages     = {54},
  year      = {2025},
  month     = {January},
  doi       = {10.1007/s10207-024-00966-1},
  issn      = {1615-5270},
}

@inproceedings{Danyil2023,
  author={Zhuravchak, Danyil and Dudykevych, Valerii},
  title = {Real-Time Ransomware Detection by Using eBPF and Natural Language Processing and Machine Learning},
  booktitle = {IEEE 5th International Conference on Advanced Information and Communication Technologies},
  pages = {1--4},
  year = {2023}
}

@article{saleh2022,
author = {Alzahrani, Saleh and Xiao, Yang and Sun, Wei},
year = {2022},
month = {01},
pages = {1-1},
title = {An Analysis of Conti Ransomware Leaked Source Codes},
volume = {PP},
journal = {IEEE Access},
doi = {10.1109/ACCESS.2022.3207757}
}

@article{gomez2018,
  author = {J. A. Gómez-Hernández and L. Álvarez-González and P. García-Teodoro},
  title = {R-Locker: Thwarting Ransomware Action Through a Honeyfile-Based Approach},
  journal = {Computers \& Security},
  volume = {73},
  pages = {389--398},
  year = {2018}
}

@inproceedings{Mehnaz2018,
  author = {S. Mehnaz and A. Mudgerikar and E. Bertino},
  title = {RWGuard: A Real-Time Detection System Against Cryptographic Ransomware},
  booktitle = {21st International Symposium on Research in Attacks, Intrusions and Defenses},
  pages = {114--136},
  year = {2018}
}

@inproceedings{Zhuravchak2021,
  author = {D. Zhuravchak and T. Ustyianovych and V. Dudykevych and B. Venny and K. Ruda},
  title = {Ransomware Prevention System Design Based on File Symbolic Linking Honeypots},
  booktitle = {11th International Conference on Intelligent Data Acquisition and Advanced Computing Systems: Technology and Applications},
  pages = {284--287},
  year = {2021}
}

@article{Lee2019,
  author = {S. Lee and H. K. Kim and K. Kim},
  title = {Ransomware Protection Using the Moving Target Defense Perspective},
  journal = {Computers \& Electrical Engineering},
  volume = {78},
  pages = {288--299},
  year = {2019}
}

@article{kok2019,
  author = {Kok, S. H. and Abdullah, Azween and Jhanjhi, NZ and Supramaniam, Mahadevan},
  title = {Prevention of Crypto-Ransomware Using a Pre-Encryption Detection Algorithm},
  journal = {Computers},
  volume = {8},
  number = {4},
  pages = {79},
  year = {2019}
}

@article{Song2016,
  author = {S. Song and B. Kim and S. Lee},
  title = {The Effective Ransomware Prevention Technique Using Process Monitoring on Android Platform},
  journal = {Mobile Information Systems},
  volume = {2016},
pages = {2946735},
  number = {1},
  year = {2016}
}

@inproceedings{Greg2018,
  author = {G. Cusack and O. Michel and E. Keller},
  title = {Machine Learning-Based Detection of Ransomware Using {SDN}},
  booktitle = {ACM International Workshop on Security in Software Defined Networks \& Network Function Virtualization},
  pages = {1--6},
  year = {2018}
}

@inproceedings{fujinoki2023,
  author = {H. Fujinoki and L. Manukonda},
  title = {Proactive Damage Prevention from Zero-Day Ransomwares},
  booktitle = {5th International Conference on Computer Communication and the Internet},
  pages = {133--141},
  year = {2023}
}

@article{Gujar2023,
title={Backup Solid State Drive for Ransomware Protection},
volume={10},
number={1},
journal={Journal of Operating Systems Development \& Trends}, 
author={Gujar, Kunal and Jagdale, Pratik and Yadav, Swapnil and Bhattacharjee, Srijita}, 
year={2023},
pages={12–18} 
}

@INPROCEEDINGS{Oujezsky2023,
  author={Oujezsky, Vaclav and Novak, Pavel and Horvath, Tomas and Holik, Martin and Jurcik, Michal},
  booktitle={46th International Conference on Telecommunications and Signal Processing}, 
  title={Data Backup System with Integrated Active Protection Against Ransomware}, 
  year={2023},
  pages={65-69},
  doi={10.1109/TSP59544.2023.10197687}}

@article{Napierone2022,
title = {NapierOne: A modern mixed file data set alternative to Govdocs1},
journal = {Forensic Science International: Digital Investigation},
volume = {40},
pages = {301330},
year = {2022},
issn = {2666-2817},
doi = {https://doi.org/10.1016/j.fsidi.2021.301330},
author = {Simon R. Davies and Richard Macfarlane and William J. Buchanan},
}

@misc{eBPF,
  author = {eBPF},
  title = {What is {eBPF}? An Introduction and Deep Dive into the {eBPF} Technology},
  howpublished = {\url{https://ebpf.io/what-is-ebpf}},
  note = {Accessed 2023-3-25}
}

@misc{bcc,
  author = {{IO Visor}},
  title = {{BCC} - Tools for {BPF}-based Linux {IO} Analysis, Networking, Monitoring, and More},
  howpublished = {\url{https://github.com/iovisor/bcc}},
  note = {Accessed 2023-3-25}
}

@article{Surati2017,
  author = {S. B. Surati and G. I. Prajapati},
  title = {A Review on Ransomware Detection \& Prevention},
  journal = {International Journal of Research and Scientific Innovation (IJRSI)},
  volume = {4},
  number = {9},
  pages = {86--91},
  year = {2017}
}

@misc{trendmicro2022,
  author = {{Trend Micro}},
  title = {Rethinking Tactics: 2022 Annual Cybersecurity Roundup},
  howpublished = {\url{https://www.trendmicro.com/vinfo/us/security/research-and-analysis/threat-reports}},
  note = {Accessed 2024-7-29}
}

@inproceedings{candar2023,
  author = {Kosuke Higuchi and Ryotaro Kobayashi},
  title = {Real-Time Defense System using eBPF for Machine Learning-Based Ransomware Detection Method},
  booktitle = {11th International Symposium on Computing and Networking Workshops},
  pages = {213--219},
  year = {2023}
}

@misc{linux_namei,
  author       = {{Linux Kernel Developers}},
  title        = {Linux kernel source: fs/namei.c},
  year         = {2026},
  howpublished = {\url{https://codebrowser.dev/linux/linux/fs/namei.c.html}},
  note         = {Accessed 2026-03-11}
}

@misc{linux_open,
  author       = {{Linux Kernel Developers}},
  title        = {Linux kernel source: fs/open.c},
  year         = {2026},
  howpublished = {\url{https://codebrowser.dev/linux/linux/fs/open.c.html}},
  note         = {Accessed 2026-03-11}
}

@misc{linux_xfs_file,
  author       = {{Linux Kernel Developers}},
  title        = {Linux kernel source: fs/xfs/xfs\_file.c},
  year         = {2026},
  howpublished = {\url{https://codebrowser.dev/linux/linux/fs/xfs/xfs_file.c.html}},
  note         = {Accessed 2026-03-11}
}

@misc{linux_ext4_file,
  author       = {{Linux Kernel Developers}},
  title        = {Linux kernel source: fs/ext4/file.c},
  year         = {2026},
  howpublished = {\url{https://codebrowser.dev/linux/linux/fs/ext4/file.c.html}},
  note         = {Accessed 2026-03-11}
}

@article{Luca2021,
title = {Kernel-level tracing for detecting stegomalware and covert channels in Linux environments},
journal = {Computer Networks},
volume = {191},
pages = {108010},
year = {2021},
issn = {1389-1286},
doi = {https://doi.org/10.1016/j.comnet.2021.108010},
url = {https://www.sciencedirect.com/science/article/pii/S1389128621001249},
author = {Luca Caviglione and Wojciech Mazurczyk and Matteo Repetto and Andreas Schaffhauser and Marco Zuppelli},
}

\end{document}